\documentclass[prb,twocolumn,epsf]{revtex4}
\usepackage{graphicx}
\usepackage{dcolumn}
\usepackage{bm}
\usepackage{hyperref}
\renewcommand{\vec}[1]{{\mathbf{#1}}}
\newcommand{\beq}{\begin{eqnarray}}
\newcommand{\eeq}{\end{eqnarray}}

\begin{document}

\title{Origin of the Mott Gap}

\author{R. G. Leigh and Philip Phillips}
\affiliation{Department of Physics,
University of Illinois
1110 W. Green Street, Urbana, IL 61801, U.S.A.}

\date{\today}

\begin{abstract}
We show exactly that the only charged excitations that exist in the
strong-coupling limit of the half-filled Hubbard model are gapped
composite excitations generated by the dynamics of the charge $2e$ boson that
appears upon explicit integration of the high-energy scale.  At every momentum, such excitations have non-zero
spectral weight at two distinct energy scales separated by the on-site
repulsion $U$.  The result is a gap in the spectrum for the composite
excitations accompanied by a discontinuous vanishing of the density of states at
the chemical potential when $U$ exceeds the bandwidth.  Consequently, we resolve the long-standing
problem of the cause of the charge gap in a half-filled band in the
absence of symmetry breaking. 
\end{abstract}

\maketitle

\section{Introduction}

In 1949, Sir Neville Mott\cite{mott} proposed that transition metal oxides with
half-filled bands possess a gap in the single-particle spectrum that
is due entirely to the energy cost for placing two electrons on the
same site.  This explanation is clearly incomplete because even in the
simplest model of a Mott insulator, the Hubbard model, none of the
eigenstates have definite local occupation.  Consequently, the charge gap in
transition metal oxides does not have the simple interpretation
as the energy gap between bands that represent electron motion among
singly and doubly-occupied
sites.   What then are the degrees of freedom that are being gapped?  Because mobile doubly
occupied sites would be inconsistent with an insulating state,
some\cite{castro,fulde} have argued that in a Mott insulator, double
occupancy is localized whereas in the metal doubly occupied sites form an itinerant fluid.
Such localization requires a dynamical degree of freedom which has not been ennunciated despite numerous simulations which display a Mott gap\cite{mg1}. In fact, the origin of
the dynamical degree of freedom that generates the elementary excitations
responsible for the Mott gap is the
essential problem of Mottness.  Knowledge of this degree of freedom
and the excitations it mediates are crucial to the physics of
high-temperature copper-oxide superconductors as they are doped Mott insulators.
Indeed, the extreme difficulty
in unearthing the 
mechanism for the localization of double occupancy led
Laughlin\cite{laughlin} to suggest that charge 
gaps in homogeneous time-resersal systems are impossible.  

In this paper, we construct explicitly the dynamical degrees of freedom that
account for the Mott gap in the absence of any symmetry breaking.
There are two key elements to our proof.  First, we show that the exact
low-energy theory for a half-filled band described by the Hubbard
model has
{\it no} bare propagating degrees of freedom of any kind. Second, we identify two dispersing
degrees of freedom or composite excitations which lead to a turn-on of
the spectral weight centered at $\pm U/2$. If $U>8t$, the composite excitations
 are orthogonal to
one another in that the spectral weight they produce never exists in the same
energy range.  The result is a gap in the spectrum.  In terms of the UV
variables, the composite excitations represent
bound states\cite{num}
involving double
occupancy or double holes and are the fundamental excitations
that define the lower and upper Hubbard bands. 

\section{Low-energy Action}

As pointed out by Laughlin\cite{laughlin}, no one has identified the
band structure of the elementary particles whose spectrum becomes
gapped at half-filling.  In this paper, we show that this can
be done by utilizing the methods\cite{lowen1} we have recently developed to
explicitly integrate out the degrees of freedom far from the chemical
potential in the Hubbard model.  We consider the Hubbard model on a
square lattice in the limit in which the bands
are well separated, that is, the on-site interaction $U$ exceeds the
bandwidth, $W=8t$, $t$ the hopping matrix element.  
At half-filling, the chemical
potential lies in the Mott gap.  As a consequence, both the degrees of
freedom above and below the chemical potential must be integrated
out if one wishes to construct a low-energy theory of the Mott insulator.  This can be done by introducing\cite{lowen1} two new fermionic fields which
when constrained appropriately will correspond to the creation of
double occupancy, $D_i$, and double holes, $\tilde D_i$.   In Lorentzian signature, the Lagrangian which makes this
integration possible,
\beq\label{hfuv}
L^{\rm hf}_{\rm UV}& =&\int d^2\theta\left[ 
iD^\dagger\dot D-i\dot{\tilde D}^\dagger\tilde D
-\frac{U}{2}(D^\dagger D-\tilde D\tilde D^\dagger)\right.\nonumber\\
&+&\left. \frac{t}{2}D^\dagger\theta b +\frac{t}{2}\bar\theta b\tilde
  D+h.c.+ s\bar\theta\varphi^\dagger (D-\theta c_\uparrow
  c_\downarrow)\right.\nonumber\\
&+&\left. \tilde s\bar\theta\tilde\varphi^\dagger (\tilde D-\theta c^\dagger_\uparrow c^\dagger_\downarrow)+h.c.\right],
\eeq
contains the two constraint charge $\pm 2e$ bosonic fields, $\varphi^\dagger_i$
(charge $2e$)
and $\tilde\varphi^\dagger_i$ (charge $-2e$) which enter the theory as
Lagrange multipliers for the creation of double occupancy and double
holes, respectively.  Mathematically, they are analogous to $\sigma$ in the
non-linear sigma model. All operators in Eq. (\ref{hfuv}) have the same site
index which is summed over. The Lagrangian also contains the integration,
$d^2\theta$, over the complex Grassman, $\theta$, and  
$b_i=\sum_j g_{ij} c_{i,\sigma}V_\sigma c_{j,-\sigma}$ is a
bond-singlet operator where
$c^\dagger_{i\sigma}$ creates a fermion on site $i$ with spin $\sigma$,
$g_{ij}=1$ iff $i$ and $j$ are nearest neighbours,
$V_\uparrow=-V_\downarrow=1$ and $s$ and $\tilde s$ are constants
appearing in the constraint which have units of energy. 
The theory (\ref{hfuv}) is completely equivalent to the Hubbard model. 
That this is so can be seen by 
integrating out $\varphi_i$ and $\tilde\varphi_i$ followed immediately
by an integation over $D_i$ and $\tilde D_i$ in the partition
function,
\beq\label{Z}
Z=\int [{\cal D}c\ {\cal D}c^\dagger\ {\cal D}D\ {\cal D}D^\dagger\ 
{\cal D}\varphi\ {\cal D}\varphi^\dagger]\exp^{-\int_0^\tau L_{\rm
    UV}^{\rm hf} dt}.
\eeq
The $\varphi$ and $\tilde \varphi$ integrations
(over the real and imaginary parts) are precisely a representation of (a
series of) $\delta$-functions of the form,
\beq
\delta\left(\int d\theta D_i-\int d\theta\ \theta c_{i,\uparrow}c_{i,\downarrow}\right),
\eeq
and 
\beq
\delta\left(\int d\theta \tilde D_i-\int d\theta\ \theta
  c^\dagger_{i,\uparrow}c^\dagger_{i,\downarrow}\right),
\eeq
We must now integrate over the $D_i$ and and $\tilde D_i$.  The
dynamical terms yield,
\beq
&&\int d^2\theta\bar\theta\theta\left[c_{i\downarrow}^\dagger c_{i\uparrow}^\dagger\partial_t(c_{i\uparrow}c_{i\downarrow})-\partial_t(c_{i\downarrow}c_{i\uparrow})c_{i\uparrow}^\dagger
  c_{i\downarrow}^\dagger\right]\nonumber\\
&&=\int d^2\theta\bar\theta\theta \sum_\sigma c^\dagger_{i\sigma} \dot
c_{i\sigma}.
\eeq
The terms proportional to $U$ lead to
\beq
&&\frac{U}{2}\int d^2\theta
\left[\bar\theta\theta c_{i\downarrow}^\dagger c_{i\uparrow}^\dagger
  c_{i\uparrow}c_{i\downarrow}-\theta\bar\theta c^\dagger_{i\uparrow}c^\dagger_{i\downarrow}c_{i\downarrow}
  c_{i\uparrow}\right]\nonumber\\
&&=U\int d^2\theta \bar\theta\theta n_{i\uparrow}n_{i\downarrow},
\eeq
the standard interaction term in the Hubbard model in the Lorentzian signature.  Finally, the
terms proportional to $V_\sigma$ yield
\beq\label{den1}
&&\int d^2\theta\ \bar\theta\theta \sum_{i,j}g_{ij}\left[c_{j,\downarrow}^\dagger 
c_{j,\uparrow}^\dagger(c_{i,\uparrow}c_{j,\downarrow}-c_{i,\downarrow}c_{j,\uparrow})\right]
+h.c.\nonumber\\
&&=\int d^2\theta\ \bar\theta\theta \sum_{i,j,\sigma}g_{ij}n_{j,-\sigma}
c^\dagger_{j,\sigma}c_{i,\sigma}+h.c.,
\eeq
after the $\varphi$ and $D_i$ integrations, whereas the
$\tilde\varphi$ and $\tilde D_i$ integrations yield the same final
result except $n_{j-\sigma}$ is replaced by $(1-n_{j-\sigma})$.
Hence, these terms add together to generate the kinetic term in the
Hubbard model.  Adding all these results together leads to 
$\int d^2\theta\bar\theta\theta {\cal L_{\rm Hubb}}= {\cal L_{\rm
    Hubb}}$, the Lagrangian for the Hubbard model.   Hence, Eq. (\ref{hfuv}) is an equivalent way of
writing the Hubbard model in which two canonical fermions describe the
physics on the $U-$scale. 

Since the physics on the $U-$scale has been
cleanly identified, the exact low-energy theory,
\beq\label{hfir}
L^{\rm hf}_{\rm IR}&=&
-\left(s\varphi^\dagger+\frac12 t b^\dagger\right)L_-^{-1}\left(s^*\varphi+\frac12 t b\right)\nonumber\\
&+&\left(\tilde s^*\tilde\varphi+\frac12 t b^\dagger\right)L_+^{-1}\left(\tilde s\tilde\varphi^\dagger+\frac 12 tb\right)\nonumber\\
&-&\left(s\varphi^\dagger-\tilde s^*\tilde \varphi\right) c_\uparrow c_\downarrow+h.c.,
\eeq
 can be constructed by
explicitly integrating out the massive fields $D_i$ and $\tilde D_i$.
Here, $L_\pm=i\frac{d}{d t}\pm \frac{U}{2}$. This integration is
straightfoward as it is strictly Gaussian.  It is important to
appreciate that the resulting theory is exact, and completely
equivalent to the Hubbard
 model. 

We would like to understand the physics of this model. It is clear from this
presentation that the bosonic and fermionic fields do not represent
weakly coupled degrees of freedom, as there are no quadratic terms in
$c_{i,\sigma}$ and the would-be boson propagators have no
poles. Presumably, one would like to find an appropriate continuum limit, but the identification of the correct continuum limit is difficult.  What we will find here is that Eq. (\ref{hfir}) is a
much better theory to work with than the Hubbard model, as it contains
the seeds of the degrees of freedom that emerge at low energies.  The
Hubbard model contains only strongly interacting electrons, and any
continuum limit that might be considered would presumably miss any
collective degrees of freedom.  

To proceed, we begin by recalling what happens in the free fermion Landau theory. There, the continuum limit is trivial to take, as we just scale towards the free fermion UV fixed point. The appropriate renormalization group flow is obtained\cite{polchinski,shankar} by scaling momenta towards the Fermi surface,  $\vec k=\vec k_F+\vec{l}$, $\vec{l}\to 0$. The fundamental reason that the latter is done is that that is {\it where the spectral density lies.} The correct degrees of freedom in the IR give rise to this spectral density. At the level of the Lagrangian,
the spectral density is determined by the vanishing of the coefficient of the quadratic terms.  The spectral density is highly peaked and the effects of renormalization are only to give weakly interacting (dressed) fermions. 

We emulate this approach by 
determining where Eq.
(\ref{hfir}) predicts the spectral density to lie.  
Fortunately, we will find that this is highly peaked (at what we will
call the upper and lower Hubbard bands), and so one might hope that the scaling towards
that locus is well-defined. We will not go so far in this paper as to claim that a continuum limit exists, but we will use this insight to establish that a dynamical Mott gap emerges. The root cause of this effect is that hidden in Eq. (\ref{hfir}) is
the dynamical degree of freedom that mediates the Mott gap: both the
electrons and bosons are locked into bound states and hence cannot
propagate independently. 
As a result the most relevant term in the Lagrangian arises from the boson-fermion interaction. 
This is a purely strongly coupled effect
which makes Mottness analogous to other problems of strong
interactions, for example QCD. An important difference with QCD which
Eq. (\ref{hfir}) lays plain is that for Mottness, the exact low-energy
Lagrangian may be derived.  This should enable an identification of the proper
collective degrees of freedom.

To proceed, we switch to frequency
and momentum space and specialize to a square lattice as our focus is
the copper-oxide plane of the cuprates.  Defining $\varphi(t) = \int
d\omega\ e^{-i\omega t}\varphi_\omega$, the energy dispersion,
$\varepsilon^{(\vec k)}_{\vec p}=4\sum_\mu\cos(k_\mu a/2)\cos(p_\mu
a)$, where $\vec k$ and $\vec p$ are the center of mass and relative
momenta of the fermion pair, and the Fourier
transform of $b_{i}$,
\beq\label{fb}
b_{\vec k}=\sum_{\vec p}\varepsilon^{(\vec k)}_{\vec p}\ c_{\vec k/2+\vec p,\uparrow}c_{\vec k/2-\vec p,\downarrow},
\eeq
we arrive at the exact working expression,
\beq\label{lfreq}
L^{\rm hf}_{\rm IR}&=&
-\frac{|s|^2}{(\omega-U/2)}\varphi_{\omega,\vec k}^\dagger\varphi_{\omega,\vec k}
+\frac{|s|^2}{(\omega+U/2)}\tilde\varphi_{-\omega,\vec k}^\dagger\tilde\varphi_{\omega,\vec k}
\nonumber\\
&+& \frac{Ut^2}{U^2-4\omega^2} b_{\omega,\vec k}^\dagger b_{\omega,\vec k}\nonumber\\
&+&(s\alpha_{\vec p}^{(\vec k)}(\omega)\varphi^\dagger_{\omega,\vec k}+\tilde s^*\tilde\alpha_{\vec p}^{(\vec k)}(\omega)\tilde\varphi_{-\omega,\vec k})\nonumber\\&\times&(c_{\vec k/2+\vec p,\uparrow}c_{\vec k/2-\vec p,\downarrow})_\omega +h.c.
\label{eq:kinterms}
\eeq
for the low-energy Lagrangian where we have suppressed the implied
integration over frequency and introduced the coupling constants,
\beq\label{eq:alpha}
\alpha_{\vec p}^{(\vec k)}(\omega)=\frac{-U+t\varepsilon_{\vec p}^{(\vec k)}+2\omega}{U-2\omega}\nonumber\\
\tilde\alpha_{\vec p}^{(\vec k)}(\omega)=\frac{U+t\varepsilon_{\vec
    p}^{(\vec k)}+2\omega}{U+2\omega}
\eeq
which play a central role in this theory.  They, in fact, will determine
the spectral weight in the lower Hubbard (LHB) and upper Hubbard (UHB) bands,
respectively. Note that in all of these expressions, $\omega$ is the
frequency of the boson field $\varphi$ or $\tilde\varphi$. As we have retained the full
frequency dependence, we will be able to determine the complete
dynamics. 

If the bosons were
weakly coupled propagating degrees of freedom, setting the coefficient
of the quadratic terms (in the Lagrangian) to zero would determine their
dispersion.  However, the coefficients of the naively quadratic terms never
vanish for any momentum and frequency. Hence, on the surface of it,
neither the bosons nor the electrons propagate and the spectral
weight vanishes at all energies. The correct theory of the Mott gap should yield, however, a
non-zero spectral weight in the UHB and LHB.  Identifying this physics
requires that we re-evaluate what should properly be considered to be
a kinetic term.  The structure of the frequency and momentum dependence of Eq. (\ref{eq:kinterms}) 
suggests that the operators
$\varphi^\dagger cc$ and $\tilde\varphi cc$ play a central role and
they determine where the spectral weight resides.  These operators
might then  be thought of as the kinetic terms for composite excitations mediated
by the charge $\pm 2e$ bosonic fields (loosely speaking, we might
think of this as occurring because of the formation of bound states).
Such an interpretation is warranted
because the spin-spin interaction and all higher-order operators contained in the $|b|^2$ term
are at least proportional to $a^4$ and hence are
all sub-dominant to the composite interaction terms.  Consequently,
at the level of the Lagrangian, the turn-on of the spectral weight is
governed by the vanishing of the coefficients of the coupled
boson-fermion terms.

That novel dynamics emerges from Eq. (\ref{eq:kinterms}) can be seen by inspection of
the coefficients (\ref{eq:alpha}). We note that the frequency poles appearing in the various terms of the Lagrangian are an artifact of our normalization, and could be absorbed into a redefinition of fields: 
$\varphi_\omega\to \sqrt{1-2\omega/U}\ \varphi_\omega$, 
$\tilde\varphi_\omega\to \sqrt{1+2\omega/U}\ \tilde\varphi_\omega$, and
$(cc)_\omega\to \sqrt{1-4\omega^2/U^2}\ (cc)_\omega$. 
 These scalings recast the Lagrangian as 
\beq
L^{\rm hf}_{\rm IR}&\to& 2\frac{|s|^2}{U}|\varphi_\omega|^2+2\frac{|\tilde s|^2}{U}|\tilde\varphi_{-\omega}|^2+ \frac{t^2}{U} |b_\omega|^2\label{eq:actlineone}\\
&&+s\gamma_{\vec p}^{(\vec k)}(\omega)\varphi^\dagger_{\omega,\vec k}c_{\vec k/2+\vec p,\omega/2+\omega',\uparrow}c_{\vec k/2-\vec p,\omega/2-\omega',\downarrow}\nonumber\\
&+&\tilde s^*\tilde\gamma_{\vec p}^{(\vec k)}(\omega)\tilde\varphi_{-\omega,\vec k}c_{\vec k/2+\vec p,\omega/2+\omega',\uparrow}c_{\vec k/2-\vec p,\omega/2-\omega',\downarrow}\nonumber\\&+&h.c.\label{eq:actlinetwo}
\eeq
Effectively, we have rescaled the coefficients $\alpha,\tilde\alpha$ to
\beq
\gamma_{\vec p}^{(\vec k)}(\omega)&=&\frac{-U+t\varepsilon_{\vec p}^{(\vec k)}+2\omega}{U}\sqrt{1+2\omega/U}\nonumber\\
\tilde\gamma_{\vec p}^{(\vec k)}(\omega)&=&\frac{U+t\varepsilon_{\vec p}^{(\vec k)}+2\omega}{U}\sqrt{1-2\omega/U},
\eeq
while the coefficients of other terms in the Lagrangian are just constants.
The first thing to notice about these expressions is that the boson frequency appears in the combinations $U\mp 2\omega$. What this will ultimately mean is that the analytic structure is concentrated around $\omega=\pm U/2$. 
\begin{figure}
\centering
\includegraphics[width=8.0cm,angle=0]{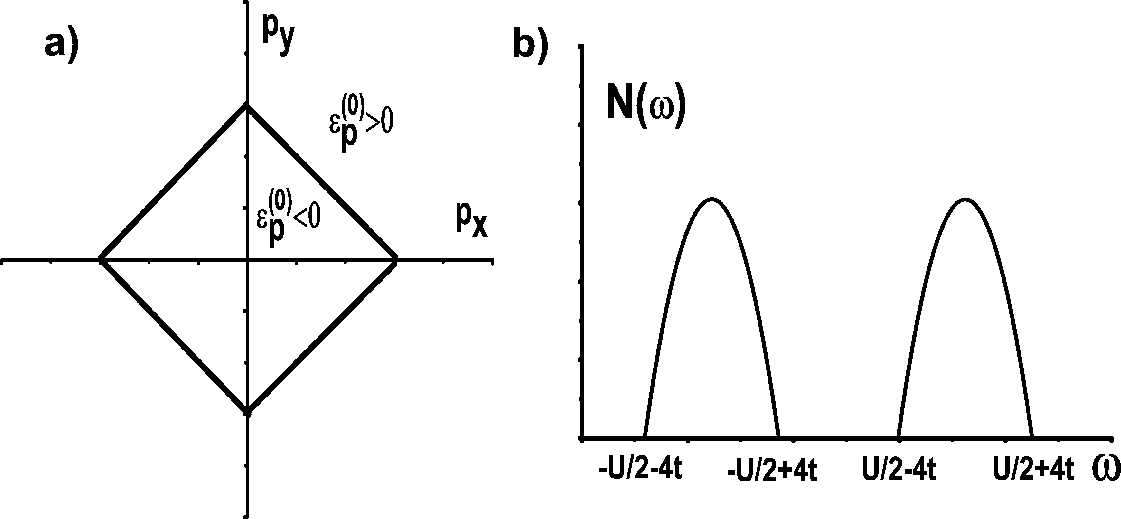}
\caption{a) Diamond-shaped surface in momentum space where the
  particle dispersion changes sign.  b) Turn-on of the spectral weight in the upper and lower Hubbard
  bands for the composite excitations as a function of energy and momentum.   In the UHB, the
  spectral density is determined to $\gamma_{\vec p}$  while for the LHB it is
  governed by $\tilde\gamma_{\vec p}$.  The corresponding operators
which describe the turn-on of the spectral weight are the composite
excitations $\varphi^\dagger cc$ (UHB) and $\tilde\varphi cc$
(LHB). The electron spectral density is determined by an overlap (see
Eq. (\ref{ov})) with
these propagating collective modes.}
\label{mottgap}
\end{figure}

\subsection{Propagating Degrees of Freedom at Half-filling: Mott Gap}

To determine where the spectral weight resides, we calculate where the
coefficients $\gamma^{\vec k}_{\vec p}$  and $\tilde\gamma^{\vec
  k}_{\vec p}$ vanish. Consider initially $\vec k=0$ so that the
dispersion simplifies to $\varepsilon^{(0)}_{\vec p}=4\sum_\mu \cos
ap_\mu$.  
When $\omega=\pm U/2$, $\gamma_{\vec p}$ ($\tilde\gamma_\vec{p}$)
vanish along the momentum surface defined by $\sum_\mu \cos p_\mu=0$.  This
surface corresponds to the diamond  $a\vec p=(ap_x,\pm\pi-ap_x)$ 
depicted in Fig. (\ref{mottgap}a).  These features define the center
of the LHB ($-U/2$) and UHB ($U/2$) for the composite excitations.
For all momenta outside the diamond, $|\vec p|>\pi$,
$\gamma_{\vec p}$ vanishes for $U/2<\omega\le U/2+4t$ while
$\tilde\gamma_{\vec p}=0$ for $-U/2<\omega<-U/2+4t$.  Within the
diamond, $|\vec p|<\pi$,  $\gamma_{\vec p}=0$ in the energy range
 $U/2-4t\le\omega <U/2$ while in the interval $[-U/2-4t,-U/2]$ the
 coefficient $\tilde\gamma_{\vec p}$ vanishes.  Consequently, for each
 momentum, spectral weight turns on at two distinct energies, one
in the
 LHB ($\tilde\gamma_{\vec p}=0$) and the other in the UHB ($\gamma_{\vec p}=0$)
 with a separation of $U$.  Note that $\gamma_{\vec p}$ and 
$\tilde\gamma_{\vec p}$ never vanish at the same energy provided that $U>8t$.
 Consequently, for $U>W$, a hard gap opens (the Mott gap) in the
 spectrum and the excitations defined by the vanishing of $\gamma$ and $\tilde\gamma$
propagate independently above and below the gap, respectively.
In terms of the composite excitations, the Mott gap opens continuously
but the spectral weight at the chemical potential rises
discontinuously as is seen in numerical calculations on
finite-dimensional lattices\cite{imada,imada2} but in contrast to the
 $d=\infty$\cite{mg1} solution.  
The composite excitations which lead to the turn-on of the spectral
weight correspond to the bound states of $\varphi^\dagger cc$ (UHB) and
$\tilde\varphi cc$ (LHB);  they represent the collective modes or in
essence the propagating charge degrees of freedom
of the half-filled Hubbard model. This is our principal conclusion.  As our analysis thus far is exact, we
conclude that in the absence of any symmetry breaking,
the coefficients $\gamma_{\vec p}$ and
$\tilde\gamma_{\vec p}$ determine the dispersion for the excitations
  that comprise the here-to-fore undefined\cite{laughlin} UHB  and
  LHB. The center-of-mass momentum $\vec k$ simply shifts the 
  momentum at which $\varepsilon_{\vec p}^{\vec k}$ changes sign, thereby keeping the
  Mott gap intact. 

Ultimately, it is the overlap between the composite excitations and
the bare electrons that
determines the 
turn-on of the electron spectral density.  Consequently, the gap in
the electron spectrum is at least that of the composite
excitations.  To determine the overlap, it is tempting to complete the
square on the $\varphi^\dagger cc$ term bringing it into a quadratic form,
$\Psi^\dagger\Psi$ with $\Psi=A\varphi+B cc$.  This would lead to
composite excitations having charge $2e$, a vanishing of the overlap
and hence no electron spectral
density of any kind.  However, the actual excitations that
underlie the operator $\varphi^\dagger cc$ correspond to a
linear combination of charge $e$ objects, $c^\dagger$ and $\varphi^\dagger c$.  In terms of the UV variables, the latter can be
thought of as a doubly occupied site bound to a hole.  To support this
claim, we construct the exact representation of the electron creation
operator at low
energies.  This can be done by adding to the starting Lagrangian a
source term
that couples to the current, $J_i$, that generates the
canonical electron operator when the constraint is solved.  In this case,
\beq
L^{\rm hf}_{\rm UV}\rightarrow L^{\rm hf}_{\rm UV}+\int d^2\theta J_{i,\sigma}\left[V_\sigma D_i^\dagger c_{i,-\sigma}\theta + V_\sigma\bar\theta c_{i,-\sigma}\tilde D_i \right] +
 h.c.\nonumber
\eeq
is the correct transformation to generate the canonical electron
operator in the UV.  If we now integrate the partition function over
$D_i$ and $\tilde{D}_i$, we find that the electron creation operator
in the IR at
half-filling
\beq\label{cop1}
c_{i,\sigma}^\dagger\rightarrow \tilde c_{i,\sigma}^\dagger&\equiv& - V_\sigma\frac{t}{U}\left(c_{i,-\sigma}b_i^\dagger + b_i^\dagger c_{i,-\sigma}\right)\nonumber\\
&+&V_\sigma\frac{2}{U}\left(s \varphi_i^\dagger + \tilde s \tilde\varphi_i\right) c_{i,-\sigma}
\eeq
is indeed a sum of two composite excitations, the first having to do
with spin fluctuations ($b^\dagger c$) and the other with
high-energy physics, $\varphi^\dagger c$ and $\tilde \varphi c$, that
is, excitations in the UHB and LHB, respectively. 
It is important to note that Eq. (\ref{cop1}) is the exact expression
for the low-energy electron at half-filling.  Consequently, we
formulate the overlap 
\beq\label{ov}
O=|\langle c^\dagger|\tilde c^\dagger\rangle\langle \tilde
c^\dagger|\Psi^\dagger\rangle|^2 P_\Psi
\eeq
for the the physical process of
passing an electron through a Mott insulator in terms of the overlap
between the bare electron with the low-energy excitations of Eq. (\ref{cop1}), $\langle c|\tilde
c\rangle$, and the overlap with the propagating degrees of freedom,
$\langle \tilde c|\Psi\rangle$ with $P_\Psi$, the 
propagator for the composite excitations.  As a result of the
dependence on the bosonic fields in Eq. (\ref{cop1}), $O$ contains 
desructive interference between states above and below the chemical potential.  Such destructive
interference between excitations across the chemical potential leads to a
vanishing of the
spectral weight at low energies\cite{imada,meinders,essler}.  Consequently, the turn-on of the
 {\it electron} spectral weight cannot be viewed simply as a sum of the spectral weight
for the composite excitations.  As a result of the destructive interference, the
gap in the electron spectrum will always exceed that for the composite
excitations.   Hence,
establishing (Fig. (\ref{mottgap})) that the composite excitations display a gap is a
sufficient condition for the existence of a charge gap in the electron
spectrum, a key conclusion of this work.

\subsection{Electron Spectral Function}

We confirm the argument in the previous section on the origin of the
gap in the electron basis by an explicit calculation of the electron
spectral function.  Because the action lacks any derivative terms 
with respect to $\varphi_i$, we can treat $\varphi$ as a
spatially homogeneous field. While {\it A priori}, such
gradient terms with respect to $\varphi_i$ are possible, their
presence at half-filling would indicate that freely propagating bosonic degree of
freedom exist at half-filling.  The absence of such terms at
half-filling makes it possible to identify that the only propagating degrees
of freedom at half-filling are gapped composite excitations. 

\begin{figure}
\centering
\includegraphics[height=8.0cm,angle=90]{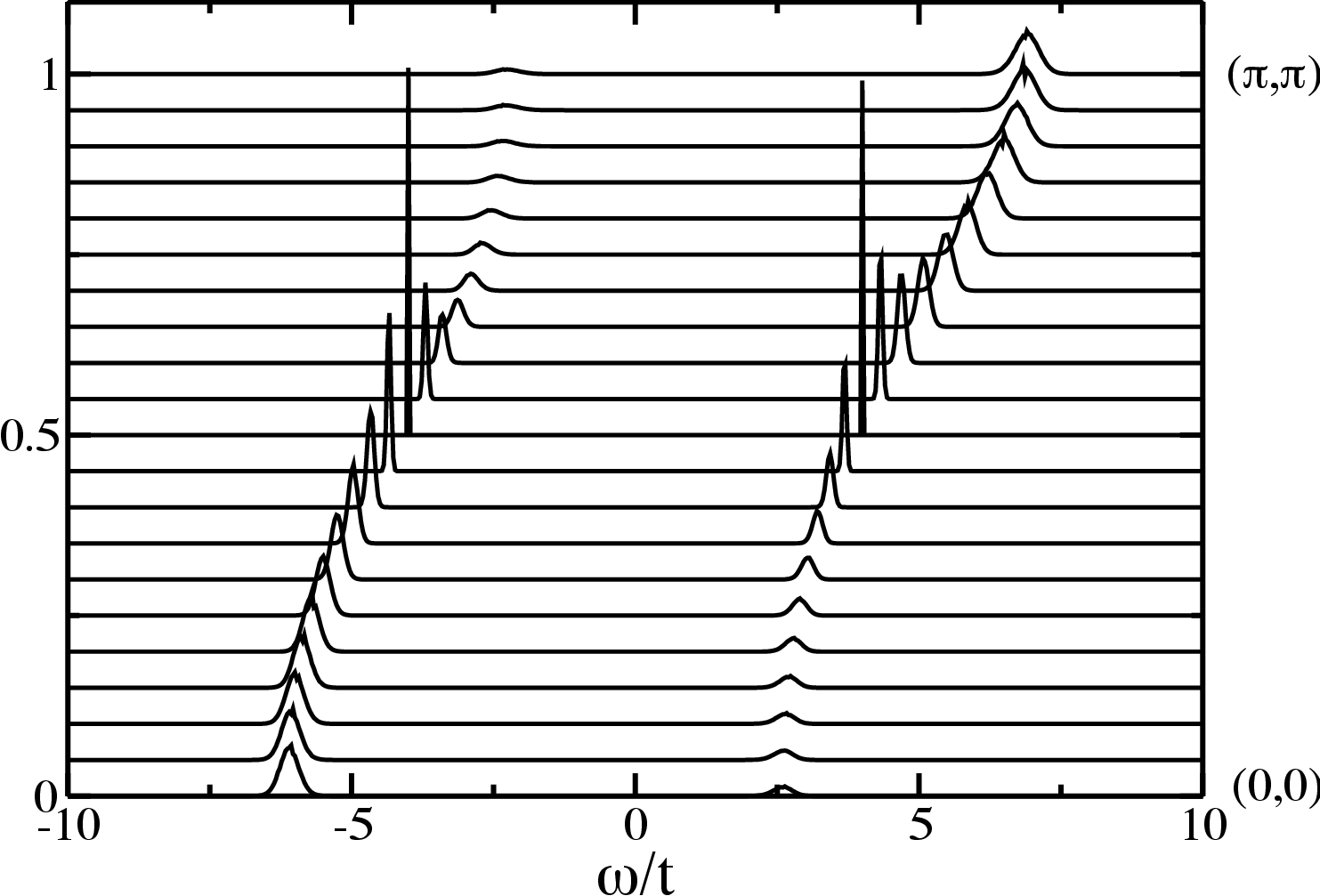}
\caption{Electron spectral function corresponding to ${\cal L}_{\rm
    Mott}$ for $U=8t$.
 The gap here is generated entirely from the dynamics of the charge
 2e bosonic fields that emerge from integrating out the upper and
 lower Hubbard bands at half-filling.}
\label{mottgap2}
\end{figure}

We proceed by rewriting the coefficient of the boson-fermi terms as
\beq
\Delta(k,\omega,\varphi,\tilde{\phi})&=&-s(\varphi^{\dagger}-\tilde{\varphi})\nonumber\\
&+&(\frac{st}{U-2\omega-i\delta}\varphi^{\dagger}+\frac{st}{U+2\omega+i\delta}\tilde{\varphi})\alpha(k)\nonumber
\eeq
and
\beq
\alpha(k)=2t(\cos(k_{x})+\cos(k_{y})).
\eeq
Any non-trivial dynamics underlying the Mott gap will arise only from the the second term in $\Delta(k,\omega,\varphi,\tilde\varphi)$.  Upon Wick rotation, $\varphi \rightarrow  i\varphi$ and 
$\varphi^{*} \rightarrow i\varphi^{*}$, we rewrite the single-particle electron Green
function as
\beq
G(k,\omega)=\int d\varphi\int
d\tilde{\varphi}G(k,\omega,\varphi,\tilde{\varphi})\exp^{-\int d\omega
{\cal  L}_{\rm Mott }}
\eeq
where ${\cal L}_{\rm Mott}$ is the IR Lagrangian with the $|b|^2$ term
dropped and
\beq
G(k,\omega,\varphi,\tilde{\varphi})=\frac{i\delta}{|\Delta(k,\omega,\varphi,\tilde{\varphi})|^{2}+i\delta}.
\eeq
As our analysis thus far demonstrates that the $|b|^2$ term has no
bearing on the Mott gap justifies our use of the truncated Lagrangian,
${\cal L}_{\rm Mott}$ which has only the charge degrees of freedom.  because of the $i\delta$ in the gap function, $\Delta(k,\omega,\varphi,\tilde{\varphi})$,
the imaginary part of the Green function
\beq\label{img}
\Im G(k,\omega,\varphi,\tilde{\varphi})&=&\lim_{\delta\rightarrow 0}\left[(U-2\omega)^{2}+\delta^{2}\right]\left[(U+2\omega)^{2}+\delta^{2}\right]\nonumber\\
&\times&\frac{\delta}{A^{2}+\left(2A(\varphi+\tilde{\varphi})+B^{2}\right)\delta^{2}+O(\delta^{4})}\nonumber\\
&=&\frac{(U-2\omega)^{2}(U+2\omega)^{2}}{B}\delta(A)
\eeq
is explicitly non-zero.  We have defined
\beq
A & = &
\left[U^{2}-4\omega^{2}-2\alpha_{k}(U+2\omega)\right]\varphi\nonumber\\
&+&\left[U^{2}-4\omega^{2}-2\alpha_{k}(U-2\omega)\right]\tilde{\varphi}\nonumber\\
B & = & 2\varphi(2\omega+\alpha_{k})+2\tilde{\varphi}(2\omega+\alpha_{k}).
\eeq
Note the arguments of the $\delta$ functions are closely related to
the coefficients $\gamma$ and $\tilde\gamma$ that led to the turn-on
of the spectral weight. However, in the electron basis, the spectral
weight is not a simple sum of the spectral weight at $\pm U/2$.  As a
result of the integration over $\varphi_i$ and $\tilde\varphi_i$, the
spectral function for the electrons arises from a complicated
interference between excitations in the LHB and UHB.  Consequently, to complete the calculation, we performed the $\varphi_i$ and
$\tilde\varphi$ integrations numerically. The resultant electron
spectral function for $U=8t$ shown in 
Fig. (\ref{mottgap2}) demonstrates clearly that a Mott gap exists and the
spectral weight is momentum dependent.  This calculation supports the
physical argument made in the previous section that the Mott gap in
the electron basis arises from a non-trivial interference between the
excitations at $\pm U/2$.   At $(\pi,\pi)$, the
spectral weight lies predominantly in the UHB whereas at $(0,0)$
it lies in the LHB.  Consequently, the real part of the Green function
must change sign\cite{dzy} along some momentum surface that lies between these
these two extreme momenta.
The location of the zero surface or Luttinger surface is the Fermi surface of the
non-interacting system as it must be\cite{zeros1,zeros2} for the half-filled system with
particle-hole symmetry.
We find then that the Mott gap arises from the dynamics of the two charge
$|2e|$ bosonic fields.  This is the first time the Mott gap has been
derived dynamically, in particular by a collective degree of freedom
of the lower and upper Hubbard bands. Relative to the gap in the
spectrum for the composite excitations that diagonalise the 
   fermion-boson terms in Eq. (\ref{eq:actlinetwo}), the gap in the electron
spectrum is larger.  This is not surprising as the bare electrons do
   not have unit overlap with the composite excitations. While our
   treatment of the charge $\pm 2e$ boson is approximate, it does
   suffice to capture the essence of of the collective mode, namely it
   mixes all sectors with varying numbers of doubly occupied sites. In
   addition, we anticipate that the electron spectral function should
   evolve as the Mott transition is approached in a similar fashion to
   that in terms of the composite particle basis.  That is, the gap
   should close continuously without a coherence peak at zero energy.
   Conseqently, the spectral weight at the chemical potential should
   jump discontinuously from zero to the value in the free system at
   the Mott transition as is seen in simulations of the Mott
   transition in finite-dimensional systems.  In
   addition, the momentum dependence of the
spectral function is identical to that obtained by dynamical mean-field
  calculations\cite{cluster1,cluster2} thereby lending crecedence to
   such cluster calculations\cite{mg1} near the Mott transition.

\section{Final Remarks}

What this analysis demonstrates is that the spin-spin interaction,
contained in the $|b|^2$ term, plays a spectator role in the generation of the Mott
gap. Nonetheless, there is
a natural candidate for the antiferromagnetic order, namely
$B_{ij}=\langle g_{ij}\varphi^\dagger_i
c_{i,\uparrow}c_{j,\downarrow}\rangle$. The vacuum
expectation value of this quantity is clearly non-zero as it is easily
obtained from a functional derivative of the partition function with
respect to $\gamma_{\vec p}$. Such an antiferromagnet, which has no
continuity with weak-coupling theory, is composed of composite
excitations which can form excitonic bound states in the two-particle
spectrum and hence is not inconsistent with the excitonic modes found
in
the mid-infrared absorption of numerous parent cuprates\cite{bk}.
 In essence, the composite excitations described by the coefficients
$\gamma_{\vec p} ^{\vec k}$ and $\tilde\gamma_{\vec p}^{\vec k}$
represent the orthogonal (they never lead to a turn-on of the spectral
weight in the same energy range) low-energy degrees modes that render the original UV problem weakly
coupled. That such new degrees of freedom emerge as the dispersing
modes is a typical feature of strong coupling.  In fact, an analogy
can be made here between our  demonstration that the propagating modes
at strong coupling in the Hubbard model are composite excitations (not
electrons) mediated by an auxiliary field that has no bare dynamics
with 't Hooft's\cite{hooft} demonstration that meson states, not free
quarks, also mediated by a non-propagating auxiliary field,  are the dispersing modes in QCD in $1+1$ dimensions. 
Our analysis suggests that a fixed point underlies the
formation of such composite excitations. Whether the $\beta-$function can be calculated within this formalism remains the outstanding question. 

\acknowledgements We thank T.P. Choy for collaboration at an early
stage of this work and for the use of his figure for the electron
spectral function and the NSF DMR-0605769 for partial support.

\end{document}